\documentclass[%
 aip,
 amsmath,amssymb,
 reprint,%
]{revtex4-1}
\usepackage{color}
\usepackage{graphicx}
\usepackage{dcolumn}
\usepackage{bm}

\usepackage[utf8]{inputenc}
\usepackage[T1]{fontenc}
\usepackage{mathptmx}
\usepackage{etoolbox}

\begin{document}

\preprint{AIP/123-QED}

\title{Modelling of Spintronic Terahertz Emitters as a function of spin generation and diffusion geometry}

\author{Yingshu~{Yang}}
\affiliation{School of Physical and Mathematical Sciences, Nanyang Technological University, Singapore, Singapore}

\author{Stefano~{Dal~Forno}}
\affiliation{School of Physical and Mathematical Sciences, Nanyang Technological University, Singapore, Singapore}

\author{Marco~{Battiato}}
\email{marco.battiato@ntu.edu.sg}
\affiliation{School of Physical and Mathematical Sciences, Nanyang Technological University, Singapore, Singapore}

\date{\today}

\begin{abstract}

Spintronic THz emitters (STE) are efficient THz sources constructed using thin heavy-metal (HM) and ferromagnetic-metal (FM) layers. 
To improve the performance of the STE, different structuring methods (trilayers, stacked bilayers) have been experimentally applied. A theoretical description of the overall THz emission process is necessary to optimize the efficiency of STE. In particular, geometry, composition, pump laser frequency,  and spin diffusion will be significant in understanding the pathways for further research developments. 
This work will apply a generalized model based on a modified Transfer Matrix Method (TMM). We will consider the spin generation and diffusion in the FM and HM layers and explain the spintronic THz emission process. This model is suitable for calculating emitted THz signal as a function of FM and HM thicknesses for different geometrical configurations. We will investigate a bilayer geometry as a test case, but the extension to a multi-layer configuration is straightforward.
We will show how the different configurations of the sample will influence the THz emission amplitude. 

\end{abstract}

\maketitle
\begin{figure}[tb]
\centering
\includegraphics[width=\linewidth]{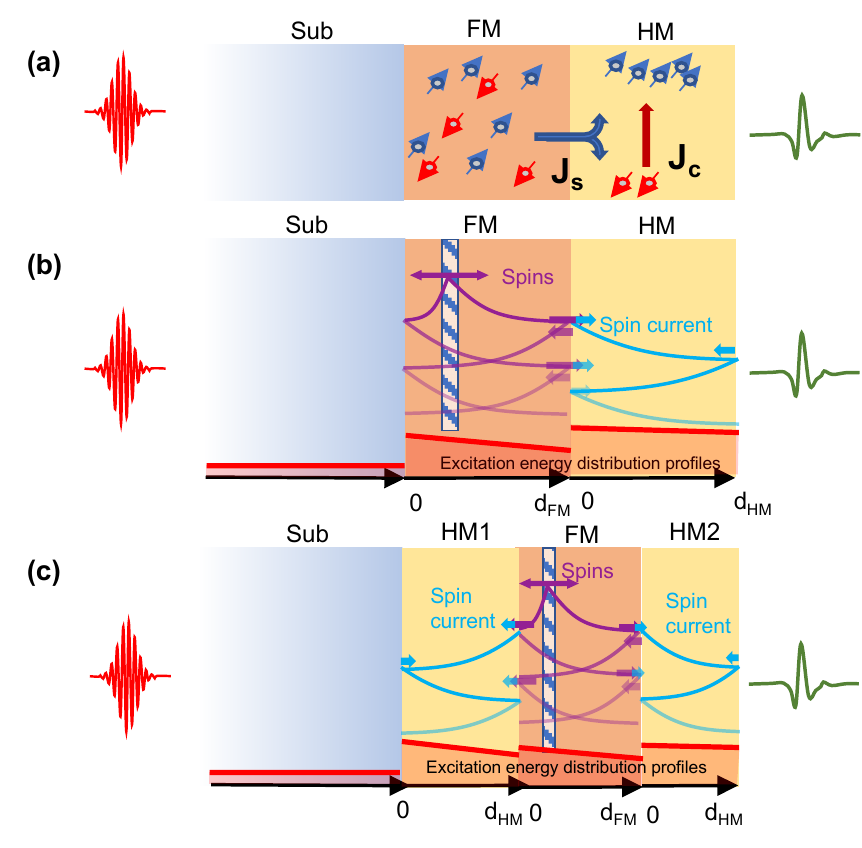}
    \caption{(a) Schematic of a spintronic THz emitter and the THz emission process. (b) and (c) Schematics of a bilayer and trilayer systems with the spin diffusion profile in the FM layer (purple line), the spin current diffusion and reflection in the HM layer (blue line), and the energy distribution profile of the laser pump along the system (red line and shaded area). }
\label{fig:Configs}
\end{figure}

THz (0.1 -30 THz) is a frequency range that is gaining popularity because of its enormous potential in basic process studies of materials as it is able to resonantly couple to conduction-electron transport, plasmons, excitons, phonons, or magnons \cite{kampfrath_resonant_2013,seifert_spintronic_2021}.
It is also an effective tool in imaging, sensing for biomedical purposes, and security applications \cite{seifert_spintronic_2021,feng_spintronic_2021,agarwal_terahertz_2022}. THz radiation can be generated using various methods, including photoconductive antennas and nonlinear optical crystals \cite{seifert_spintronic_2021}.
Spintronics THz emitters (STE), a new type of THz source, based on spin-to-charge conversion, have gained popularity in recent years due to their high efficiency, broad bandwidth, and ease of manufacture.
STE is usually constructed using two simple layers of ferromagnetic-metal(FM), and heavy-metal(HM)\cite{kampfrath_terahertz_2013,seifert_efficient_2016,wu_high-performance_2017}.
The mechanism at the core of THz emission first involves the photoexcitation of the FM layer.
This induces a spin population that diffuses to the HM layer in a process called superdiffusive spin transport \cite{battiato_superdiffusive_2010,battiato_theory_2012,battiato_treating_2014,battiato_ultrafast_2016,battiato_spin_2017,carva_ab_2011,carva_ab_2013}.
Here, due to the high spin-orbit-coupling (SOC) of the HM, inverse Spin Hall effect (ISHE) \cite{seifert_efficient_2016} will take place and produce a transverse charge current responsible for the THz emission (as shown in Fig.~\ref{fig:Configs}(a)).

The study of STE optimization has recently received much attention in order to broaden their applications. As a result, there have been different experimental studies carried out to test the THz generation efficiency based on different materials of the structure\cite{seifert_efficient_2016,seifert_terahertz_2018,qiu_layer_nodate,liu_spintronic_2022}, different thicknesses of the layers\cite{zhou_broadband_2018,torosyan_optimized_2018,qiu_layer_nodate,wu_high-performance_2017}, and even different stacks of layers\cite{feng_highly_2018,cheng_studying_2021,seifert_efficient_2016,cheng_far_2019,herapath_impact_2019}. 
To analyze the performance of the STE based on the above approaches, a theoretical model to describe the STE geometry and THz emission process is needed. Models based on basic THz emission processes involving the pump-pulse absorption, spin generation, and spin to charge conversion have been developed and used in different analysis \cite{seifert_efficient_2016,torosyan_optimized_2018}. These models usually calculate the THz emission amplitude assuming the spin current amplitude to be proportional to the excitation energy deposition \cite{seifert_spintronic_2021} and at the same time assuming the FM/HM interface is transparent to spins and electrons \cite{seifert_spintronic_2021,seifert_efficient_2016}.
However, no general model has considered the detailed geometry of the excitation laser and spin current with specific STE structures in the THz emission process.

In a typical experimental setup, STEs are constructed by combining substrate, FM, and HM layers in a number of different arrangements. The ordering of the layers, their thicknesses, and the side illuminated by the laser pump will impact the geometry of the spin generation, the spin diffusion, and the energy profile of the laser pump.
The model we applied here addresses all of the above properties on the same footing.
Specifically, we want to describe the three fundamental processes that characterize any STE: the propagation of the laser excitation pump through the system, the generation and diffusion of spins in the HM and FM layers, and the generation and propagation of the THz pulse throughout the system.
These processes are controlled by a number of parameters, namely, the type of materials, their thicknesses, the arrangement of the layers, the frequency of the pump, the spin diffusion length, and the degree of spin reflections between interfaces. Fig.\ref{fig:Configs}(b) and (c) show schematics of the processes mentioned above in both bilayer and trilayer systems.
The dependence of the outgoing THz pulse strongly depends on the choice of these parameters.
Our findings show that properly fitting the material properties in optical and THz regions is necessary to achieve quantitative predictions.

\section{Methods}

To properly model STEs, we have to describe the three sub-processes that occur in the THz emission: (1) Propagation and absorption of the optical pump into the STE's layers; (2) Production of the spin current, its propagation through the STE's layers and the eventual conversion into a transversal charge current; and (3) Production, propagation and extraction from the STE of the THz electromagnetic radiation. The following sections will address each of these tasks individually.

\subsection{Pump laser absorption profile}\label{sec:Absorption}

We describe the optical EM waves propagation and absorption using a Transfer Matrix Method (TMM) \cite{coutaz_principles_2018,born2013principles}.The system is composed by three layers: a Substrate, a FM layer and a HM layer, in different stacking orders. According to the standard Transfer Matrix Method, the transmission and reflection of the waves throughout an layer sample can be expressed as,
\begin{equation} \label{eq:TMM1}
	\begin{bmatrix} f_{\infty}^>\\ f_{\infty}^<\end{bmatrix} = \bar{\bar{T}}_{[0,\infty]} \begin{bmatrix} f_{0}^>\\  f_{0}^<\end{bmatrix},
\end{equation}
where $ \bar{\bar{T}}_{[0,\infty]}$ is the frequency-dependent $2\times 2$ transfer matrix that propagates the fields from the beginning to the end of the multilayer, and $f_{0}$, $f_{\infty}$ represent the field amplitudes at the beginning and the end of the multilayer. The superscripts $^>$ and $^<$ represent the right and left propagating waves respectively.
Assuming the pump pulse is impinging on the sample from the left, $f_{0}^{>}$ is the time profile of the pump (which we assume known). Generally, there will be no second pump incoming from the right, so $f_{\infty}^{<} = 0$. The two remaining field amplitudes $f_{\infty}^{>}$ and $f_{0}^{<}$, which represent the transmitted and reflected waves respectively, are the unknowns in the system in Eq.~\ref{eq:TMM1}. We remind that the system of two equations in Eq.~\ref{eq:TMM1} is to be solved for every frequency independently.

The generation of the spin currents and the subsequent diffusion of the spins in the FM and HM layers strongly depends on how the energy deposited by the laser pump is partitioned through the system \cite{zhou_broadband_2018,torosyan_optimized_2018}.
According to the Poynting theorem, the total energy loss due to Joule effects can be expressed as, 
\begin{equation}\label{eq:heat}
    Q_{loss} = -\int _{-\infty}^{+\infty} \mathrm{d}t \oint _{S} ( \mathbf {E} \times \mathbf {H}) \cdot \mathrm{d}  \boldsymbol{S}.
\end{equation}
Here, $Q_{loss}$ is the total energy dissipated by a system enclosed by the closed surface $S$.
Because of the planar symmetry of STE, the surface $S$ can be chosen to be a parallelepiped enclosing the layer. At normal incidence, only faces of $S$ parallel to the interfaces between layers contribute to the integral in Eq.~\ref{eq:heat}. Hence, we can define the energy per unit area that crossed a surface $S_z$ at position $z$ as,
\begin{equation}\label{eq:flux}
    \Phi(z) = - \int_{-\infty}^{+\infty} E[t,z] H[t,z] \mathrm{d}t,
\end{equation}
where $z$ is the position of the surface $S_z$.
Thus, if we assume the two interface position of a layer with thickness $d$ in a multilayer system as $z_0$ and $z_0+d$, and assuming the initial input of the pump laser as $Q_{in}$, the net absorption for this layer will be: 
\begin{equation}\label{eq:absorb}
 A_{layer} = \frac{Q_{loss}}{Q_{in}} =  \frac{\Phi(z_0) -  \Phi(z_0 + d)}{Q_{in}}.
\end{equation}
where $Q_{in}$ can also be calculated using Eq.~\ref{eq:flux} with fields at the initial surface of the system by ignoring the reflected fields ($f_0^<$ in Eq.~\ref{eq:TMM1}). This net absorption can be used when simulating STEs with fixed FM thicknesses. Similar ideas can be used to obtain the energy distribution profile. If we assume a local axis within one single layer, the energy distribution profile can be expressed as,
\begin{equation}\label{eq:energy_profile}
D(z) = -\frac{d (\Phi(z)/Q_{in})}{dz}
\end{equation}
The energy distribution profile of Eq.\ref{eq:energy_profile} is schematically displayed at the bottom of Figs.~\ref{fig:Configs}(b) and (c) with a red solid line.

\subsection{Spin generation in the FM layer}

We now address how the absorbed energy in the FM layer produces a spin current profile over the whole sample. We focus first on how energy deposited between $z$ and $z+dz$ within the FM layer propagates through the whole sample. We assume that the spin current spatial distribution generated by an infinitesimally thin layer and in the absence of interfaces is in the form of an exponential decay, with an effective spin diffusion length $\lambda$, both towards the left and the right of the emission plane (See Fig.~\ref{fig:Configs}). However due to the finite thickness of the FM layer, as well as the presence of the other layers, we must explicitly consider the reflections and the transmissions of the spins at the layers' boundaries.

We call $\alpha$ and $\beta$ the probabilities for a spin to be transmitted over interfaces to the right and the left of the considered spin current emission point $z$. Such probabilities depend on the materials and the quality of the interfaces. We assume that the probability of transmission at the interface with air or with the substrate (assumed insulating, with sufficiently large bandgap to prevent injection of the excited spin currents) to be 0. The rate at which the spin population is transmitted to the left or to the right can be calculated by exploiting the properties of geometric series (see Suppl. Info. for the detailed derivation).
By summing up the contributions of all the multiple reflections at the boundaries, we obtain the following expressions
\begin{align}
\label{eq:zetaR}
\zeta_R(z) &= \frac{\alpha}{N}  \left[ e^{-\frac{d-z}{\lambda}} + \bar{\beta} e^{-\frac{d+z}{\lambda}}\right], \\
\label{eq:zetaL}
\zeta_L(z) &= \frac{\beta}{N}  \left[ e^{-\frac{z}{\lambda}} + \bar{\alpha} e^{-\frac{2d-z}{\lambda}} \right],
\end{align}
where $N={1-\bar{\alpha} \bar{\beta} e^{-\frac{2d}{\lambda}}}$, $\bar{\alpha} = 1-\alpha$, $\bar{\beta} = 1-\beta$.
Eqs.~\ref{eq:zetaR} and \ref{eq:zetaL} measure the number of spins (generated at position $z$) that have crossed (transmitted) the right or left interfaces of the FM layer, respectively. We call $\zeta(z)$ the spin generation efficiency profile function (purple solid lines in Fig.~\ref{fig:Configs}).

Finally, to compute the effective number of spins that is injected from FM to HM, we must consider the effects of the laser pump.
We assume that the spin density is proportional to the laser fluence at the point $z$ times the left or right spin generation efficiency profile.
The total spin population becomes
\begin{equation}
\label{eq:totalspin}
    S(d) = \int_{z_i}^{z_i+d} D(z) \zeta(z) dz,
\end{equation}
where $D(z)$ is the energy (per unit area) deposited by the laser pump (Eq.~\ref{eq:energy_profile}), $z_i$ is the position of the FM layer and $\zeta(z)$ must be chosen either to be right or left, depending on the geometrical configuration of the system.
For example, if we consider the bilayer THz emitter shown in Fig.\ref{fig:Configs}(b), we need to calculate only the transmittance to the right interface. However, if we want to model the trilayer case shown in Fig.\ref{fig:Configs}(c), then the transmittance for both the left and right interfaces have to be considered.

\subsection{Spin current profile in the HM layer}
After the laser pump has excited the spin population in the FM layer, spins start diffusing and are eventually injected at the edge of the HM layer.
We again assume an exponentially decaying spatial distribution with an effective spin diffusion length $\lambda$.
We call $\gamma$ and $\mu$ the probabilities for a spin to be transmitted over the right and the left interfaces, respectively.
We want to calculate the average number of spins at a given position $z$ inside the HM layer.
Again, by exploiting the properties of geometric series (see Supp. Info.)\cite{Yang_Secondary_2022}, we can sum up the contributions of the multiple reflections and obtain the following
\begin{equation}\label{eq:HMdiff}
\sigma(z, d) = \frac{1}{N} \left[ e^{-\frac{z}{\lambda}} + \bar{\gamma} e^{-\frac{2d-z}{\lambda}} \right],
\end{equation}
where $N = 1- \bar{\gamma} \bar{\mu} e^{-\frac{2d}{\lambda}}$, $d$ is the thickness of the HM layer and $\bar{\gamma} = 1-\gamma$, $\bar{\mu} = 1-\mu$.
The above equation measures the spin current density at position $z$ for a given HM layer of thickness $d$ (see Fig.~\ref{fig:Configs} light blue curves). 
The transversal charge current density is finally obtained by multiplying by the inverse spin Hall coefficient of the HM, similarly to our previous work. \cite{yang_removal_2021,yang_transfer-matrix_2021}

Two other mechanisms of generation of transversal charge curret are also possible.
The first one is caused by spin-to-charge current conversion in the FM layer. This contribution is usually much smaller than the HM layer contribution due to the low spin to charge conversion efficiency of FM layer\cite{seifert_spintronic_2021} and will be neglected in this work.
A second mechanisms involves the creation of hot electrons in the HM layer. These hot electron will diffuse back to the FM layer and act as a secondary excitation of the FM layer.
This enhancement will become large when HM layer is thick \cite{Yang_Secondary_2022}.
We stress that the above contributions, even if ignored in most cases, can be described by our model (but it is not included in this work for simplicity). The FM contribution can be included by adding a source layer to the modified TMM model developed in Sec.\ref{sec:TMM}. The secondary enhancement from the HM layer can also be added by specifically calculating the absorption of the pump-pulse in the HM layer, which can be calculated using Sec.\ref{sec:Absorption}. 
As we are more interested in the performance of the main contribution, we show results of the HM layer emission only in the following section. We stress that the full model of including all three contributions is straightforward. However, specifying the percentage of each contribution in experiment requires detailed material data fitting (both THz frequencies and pump laser frequencies) to increase the accuracy.

\subsection{THz radiation production and propagation}\label{sec:TMM}

The final step is  computing the THz radiation extracted from the STE. This requires the computation of the production of the THz within the HM layer as well as its propagation though the multilayer. In this case, standard TMM cannot be used. In the case one of the layers act as a source of electromagnetic radiation (by mean of a time and position dependent volume current) we use the modified TMM, which we call TMM-with-source, that we developed in Ref.~\onlinecite{yang_transfer-matrix_2021}. The expression that we obtained maintains the structure of the TMM, but includes a source term 
\begin{equation} \label{eq:EQonlyemission}
	\begin{bmatrix} f_{\infty}^>\\ f_{\infty}^<\end{bmatrix} = \bar{\bar{T}}_{[0,\infty]} \begin{bmatrix} f_{0}^>\\  f_{0}^<\end{bmatrix} + \begin{bmatrix} J^> \\ J^<\end{bmatrix} .
\end{equation}
where $J^>$ and $J^<$ are the amplitudes of the right and left propagating fields generated in the source layer and account for time and position dependent charge currents within the multilayer. In this work, the position dependent profile used for the construction of the source term is taken as Eq.~\ref{eq:HMdiff} and the time dependent profile is taken as a similar shape as used in Ref.~\onlinecite{yang_transfer-matrix_2021}.

It should be noticed that, while in the case of TMM applied to the absorption of the optical pump laser, in the case of TMM-with-source for the emitted THz radiation the constraints on the field amplitudes on the right and left of the sample are different. In this case, no external THz pulse is sent to the multilayer, and therefore $f_{\infty}^<$ and $f_{0}^>$ are to be set to zero. This means that in a THz emission process the only source is the THz generated from the HM layer and Eq.~\ref{eq:EQonlyemission} can be solved for the $f_{\infty}^{>}$ and $f_{0}^{<}$ amplitudes.  To be noticed that if we want to describe a trilayer system with two HM layers acting as sources (case shown in Fig.\ref{fig:Configs}(c)), the calculation will be straightforward with two additional terms describing fields coming from two different layers. 

\begin{figure*}[tb]
\centering
\includegraphics{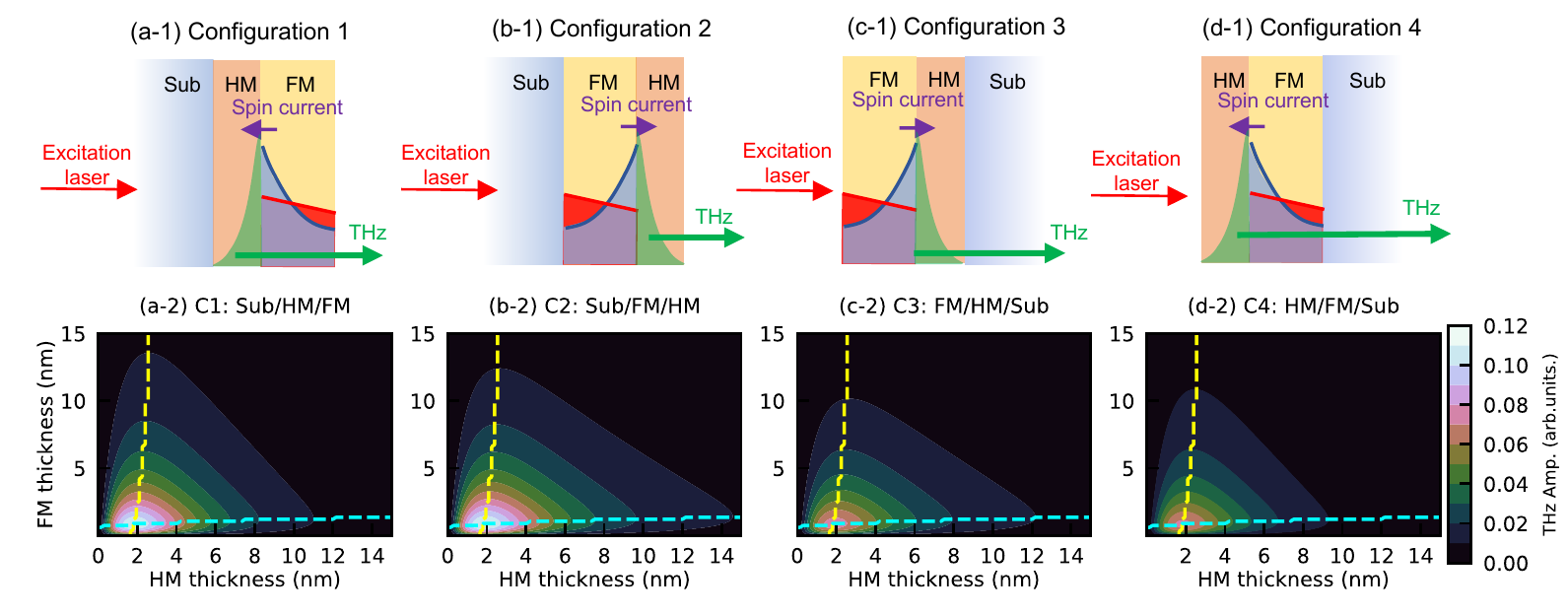}
\caption{The schematic of the four different configurations of STE and its corresponding emission map. (a-1)--(d-1) Different configurations of STE structure and a schematic show of the energy distribution profile (red shaded, Eq.\ref{eq:energy_profile}), overall spin generation efficiency profile (blue shaded, Eq.\ref{eq:zetaL}), and spin current diffusion profile (green shaded, Eq.\ref{eq:HMdiff}) for each configuration. (a-2)--(d-2)The THz emission map with changing Pt(HM) (0-15nm) and Co(FM)(0-15nm) thicknesses for four different configurations.The yellow dashed line represents the peak position of the THz emission profile as a function of HM thickness, while the blue dashed line represents the peak position of the THz emission profile as a function of HM thickness. }
\label{fig:CC_EM}
\end{figure*}

\begin{figure*}[tb]
\includegraphics{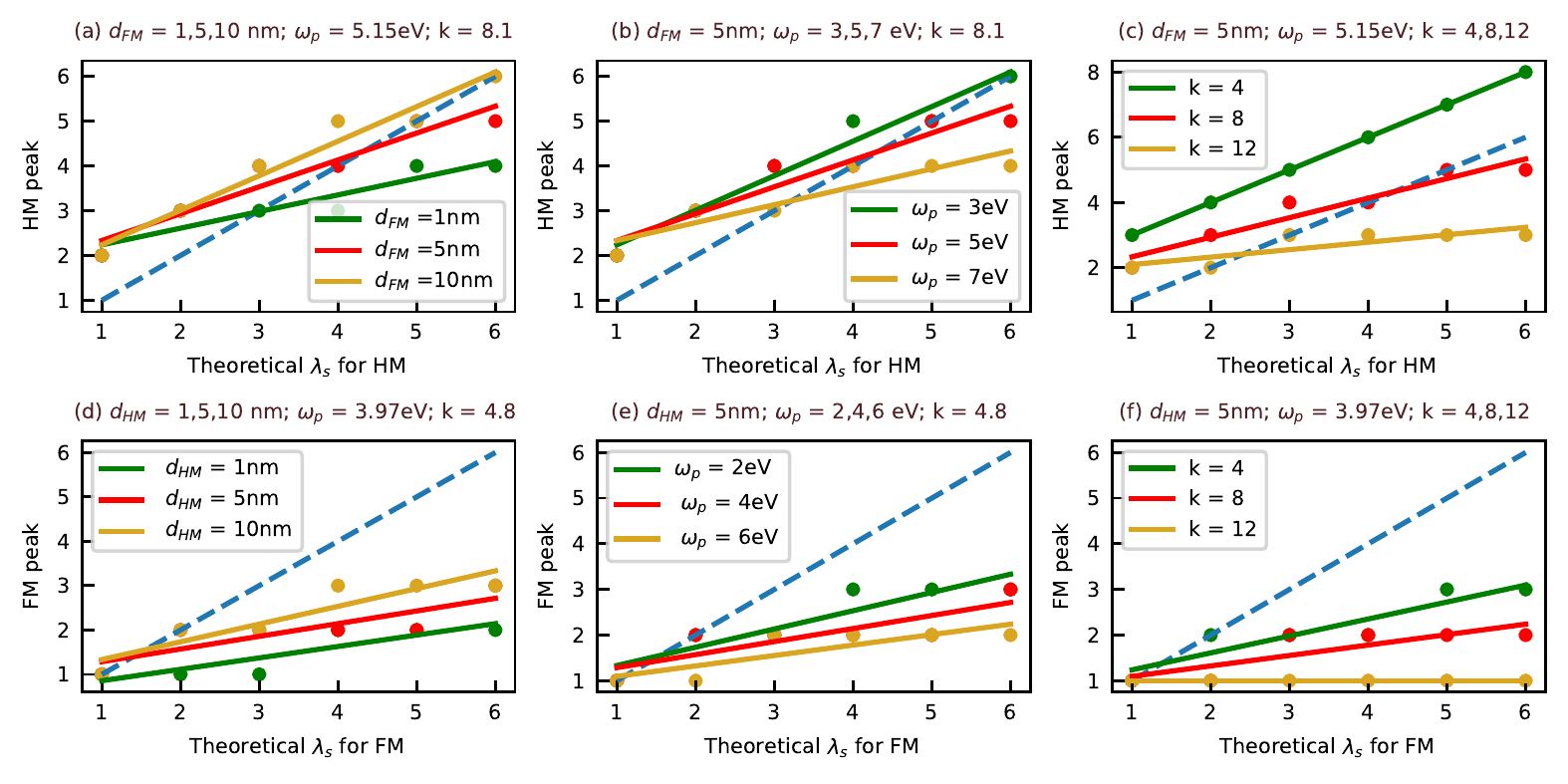}
\caption{ The linear fit of THz emission profile peak positions with changing theoretical Spin diffusion length of FM and HM under different conditions. The first row, (a)-(c), shows the peak position change of the THz emission profile as a function of HM thickness when the theoretical spin diffusion length for the HM layer changes from 1-6nm with the plasma frequency of HM layer for THz range material properties is changing from 3-7eV, and the extinction coefficient for optical range material properties is changing from 4-12. The second row, (d)-(f), shows the peak position change of the THz emission profile as a function of FM thickness when the theoretical spin diffusion length for the FM layer is changing from 1-6nm with the plasma frequency of FM layer for THz range material properties is changing from 3-7eV, and the extinction coefficient for optical range material properties is changing from 4-12.}

\label{fig:All_cases}
\end{figure*}

\section{Results}

In the following, we present the calculated THz amplitudes as a function of geometry, layer arrangements, materials, and laser pump frequency.
In Fig.~\ref{fig:CC_EM}(a-1)--(d-1), we show the four different layer arrangements we considered in this work and label them C1, C2, C3, and C4, respectively.

We start with a simple STE structure constructed with quartz (substrate, 1mm), Co (FM), and Pt (HM, 0--15nm).
The material properties of Co and Pt at THz frequencies are taken from Ref.~\onlinecite{ordal_optical_1985}, while at optical frequencies taken from Ref.~\onlinecite{werner_optical_2009}. The optical properties for quartz are taken from Ref.~\onlinecite{Gao_thin_2012} and at the THz range are taken as the experimentally measured dielectric constants in Ref.~\onlinecite{Yang_Secondary_2022}.
We take the Pt and Co spin diffusion length to be 1.1 nm and 1nm, respectively,\cite{seifert_efficient_2016,torosyan_optimized_2018,zhou_broadband_2018,zhang_determination_2013} as a test case.

Figs.~\ref{fig:CC_EM}(a-2)-(d-2) show the calculated THz emission peak amplitude for different HM and FM thicknesses for the C1, C2, C3, and C4 configurations. One can notice that at parity of FM thickness, the emitted THz intensity increases at first, as more HM thickness allows for more efficient spin-to-charge conversion. However, eventually, further increases do not provide further gain in the THz intensity but become detrimental as larger metallic regions lead to absorption of the generated THz radiation within the sample itself. This reproduces the known fact that the THz intensity peaks for relatively thin layers for both changing HM thickness and FM thickness.\cite{seifert_efficient_2016,seifert_spintronic_2021,cheng_studying_2021,torosyan_optimized_2018}

The THz emission amplitude depends on the arrangement of the layers. We observe that exciting from the substrate side produces stronger THz emission compared to exiting from the active bilayer side (C1 and C2 $>$ C3 and C4). The reason is that in the second cases, the produces THz radiation has to traverse the quartz substrate, which absorbs in that frequency range. This reproduces experimental findings on quartz substrates \cite{Yang_Secondary_2022}. The situation is reversed in the case of a sapphire substrate (material properties taken from Ref.~\onlinecite{sanjuan_optical_2012} for THz range and Ref.~\onlinecite{Kelly_Program_1972} for optical range), where instead absorption of the pumping radiation in the substrate becomes more important. Experiments confirm this scenario.\cite{liu_separation_2021}

Apart from intensity, we can also extract the behaviour of the emitted THz with changing layers' thicknesses. Although configurations C1 and C2 display similar maximum THz intensities, their behaviour with layers' thicknesses is different. We can observe that the most relevant characteristic controlling these dependences is which one of the two active layers (HM or FM) faces the pumping laser (see Fig.~\ref{fig:CC_EM}). Each layer has three key effects in the THz production process. The FM acts as the generator of spin current, as absorber of optical photons, and absorber of THz radiation, while the HM as spin-to-charge converter and, again, as absorber of both optical and THz frequencies. The thickness dependence of all this processes depends on the material properties. Yet the relative weights of each effect in each layer depends on the relative positions. For instance the role of optical photons absorber is higher in the first layer traversed by the pump pulse. 

One further interesting finding is that the peak positions of the THz emission with changing FM (or HM) layer thickness are not fixed (yellow dashed line for HM peak position, blue dashed line for FM peak position) and they depend on the thickness of the adjacent layer. However, the optimal HM thickness at which the THz emission is maximal is generally used in experiments as a quick and quantitative estimation of the spin diffusion length and vice versa\cite{papaioannou_thz_2021,seifert_spintronic_2021,zhou_broadband_2018,torosyan_optimized_2018,seifert_efficient_2016}. For that to be a meaningful estimation, the peak position should only depend on the spin diffusion length and not be affected by other characteristics of the sample. However, this contradicts our findings. 

To understand this better, we performed a larger set of calculations where we compare how those peak positions compare with the actual spin diffusion length when other parameters are changed. In Figs.~\ref{fig:All_cases}(a)-(c) we show on the y axis the thickness of the HM layer for which we obtain the strongest THz emission, while on the x coordinates the spin diffusion length in the HM used in the calculations. To really claim that the peak HM thickness can be used as an estimation of the spin diffusion length, one should require  the points to be over, or close enough to, the $y=x$ line (blue dashed line in Figs), or more generally have a stable functional dependence unaffected by other layer's properties. However, while a sufficiently linear relationship can be found between the two quantities, important deviations can be observed from the desired correlation. 

Yet even more crucially, the relationship between the two quantities is very strongly dependent on the other parameters of the system. In Fig.~\ref{fig:All_cases}(a) the thickness of the FM layer is shown to impact the peak position. For instance if the measured HM peak thickness were $4$nm, the extrapolated spin diffusion should be $3$nm if the FM is $10$nm thick, or twice as large for a sample with a $1$nm thick FM. One could still argue that the FM thickness is generally known and one could couple experimental results to theory to do more reliable estimations of the spin diffusion length. However Figs.~\ref{fig:All_cases}(b) and (c) show that that will require a very careful characterisation of the sample. A change in the dielectric properties of the HM (THz plasma frequency $\omega_p$ and optical extinction coefficient $k$) can, in fact, strongly impact the relationship (in particular Fig.~\ref{fig:All_cases}(c) shows the huge change in the dependence, with increasing extinction coefficient).

Similarly, we compute the FM layer thickness at which one obtains the highest THz emission at parity of HM thickness. It might be tempting to use that to estimate the spin current diffusion length in the FM. However Figs.~\ref{fig:All_cases}(d)-(f) again show that the correlation between the two quantities is far too strongly dependent on other properties of the multilayer.

\begin{figure}[tb]
\centering
\includegraphics[width=\linewidth]{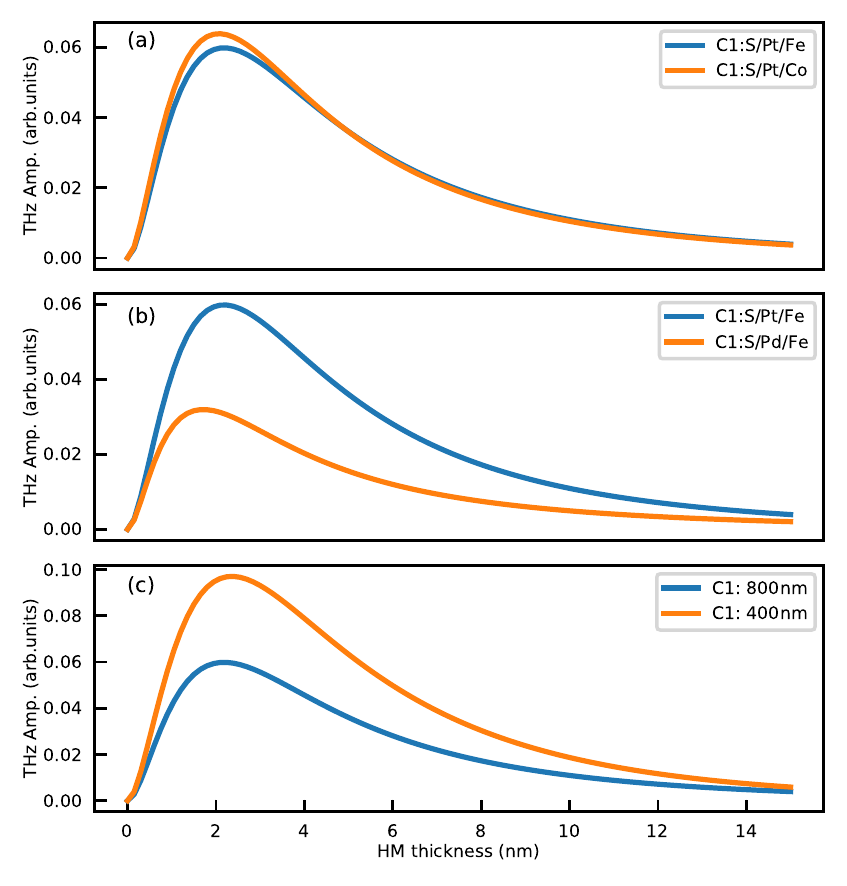}
\caption{Three different tests of THz emission. (a) THz emission profile for the same Sub/HM/FM configuration with different FM material. (b) THz emission profile for the same Sub/HM/FM configuration with different HM material. (c) THz emission profile for the same Sub/Pt/Fe configuration with different excitation lasers. }
\label{fig:CaseCompare}
\end{figure}

We now move onto analysing the dependence of the THz emission on the used materials and pump laser frequency. We  choose configuration C1 as the test case and describe three material combinations.
We set the HM material as Pt and Pd and the FM material as Co and Fe (material properties for pump laser frequencies  taken from Ref.~\onlinecite{werner_optical_2009} and for THz frequencies  from Ref.~\onlinecite{ordal_optical_1985}.)
We set the FM thicknesses to 3nm, the spin diffusion length $\lambda=1.1$nm for the HM, and $\lambda=1$nm for the FM cases.
Fig.~\ref{fig:CaseCompare}(a) and (b) show the calculated THz amplitudes. 
Changes are direct consequence of the different optical properties of the materials.
Fig.~\ref{fig:CaseCompare}(c) shows the dependence of the THz pulse for two different pump laser wavelengths (400nm and 800nm). We see that, for the given thicknesses and $\lambda$, the THz emission profile changes. Specifically, the 400nm emitted THz amplitude is higher than 800nm one. This is in qualitative agreement with experiments \cite{adam_magnetically_2019}.

\begin{figure}[tb]
\centering
\includegraphics[width=\linewidth]{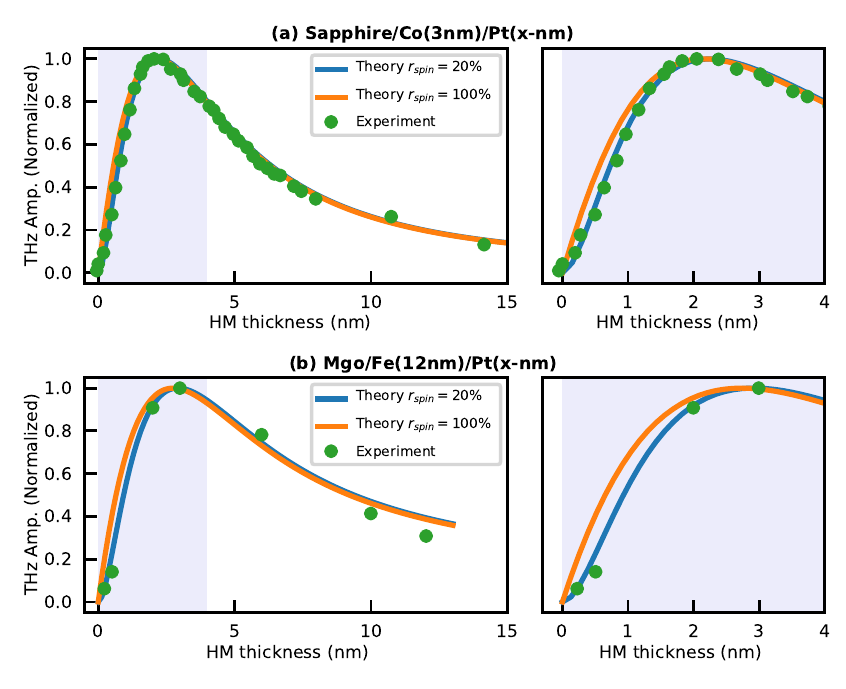}
\caption{The comparison of THz emission profile with changing thicknesses of HM between experimental data and theoretical calculation when spin reflection at FM/HM interface is 20\% and 100\%. (a) Comparison for sapphire/Co/Pt structure with data taken from \cite{zhou_broadband_2018}. (b) Comparison for MgO/Fe/Pt structure with data taken from \cite{torosyan_optimized_2018}.}
\label{fig:Spinloss}
\end{figure}

Finally, we show another interesting finding for the THz emission profile at low HM layer thicknesses. From the generalized spin diffusion model (Eq.~\ref{eq:HMdiff}) in the HM layer, we know that the spin reflections at the interfaces will play a more significant role at lower compared to higher thicknesses. To see the influence of the spin reflections, we calculated two sets of THz emission profiles as a function of HM layer thicknesses. Then, we compared them to two different sets of normalized experimental data for Sapphire(1mm)/Co(3nm)/Pt(xnm) and MgO(0.5mm)/Fe(12nm)/Pt(xnm) samples taken from Refs.~\onlinecite{torosyan_optimized_2018,zhou_broadband_2018}.

In these two sets of calculations, we took the spin reflection percentage as 20\% and 100\% for the FM/HM interface and compared them to the experimental data ($\bar{\mu} = 20\%$ and $\bar{\mu} = 100\%$ in Eq.~\ref{eq:HMdiff}). We can see that when 20\% of spins are reflected (80\% spins transmitted), the profile shows a much better fit to the experimental data at lower thicknesses. This can be evidence to show that one possible reason for the appearance of a positive second derivative at low thicknesses in the experiment comes from the spin reflections at the interfaces in the HM layer. Hence, it is crucial to consider the spin reflections when dealing with low HM thicknesses.

\section{Conclusion}

We built a theoretical model that included a spin diffusion profile in HM, a spin generation efficiency profile in FM, and an excitation laser energy profile in FM with the modified TMM to describe the spintronic THz emission.
The thicknesses of the layers and the substrate, the material choice, the layer arrangement, and the pump laser frequency all affect the THz emission profile.
Using this model, we showed that the peak position of the THz pulse as a function of the HM thickness is dominated by the spin diffusion length. However, other contributions, such as laser absorption and layer arrangements, play a non-negligible role.
We also showed that an accurate description of the spin current reflection at interfaces is important to achieve accurate predictions at low thicknesses ($d_{HM}< 4nm$). This is important because the THz signal usually peaks within this thickness range .
Finally, we showed that a proper experimental fitting needs to be done to extract reliable information about the STE, such as spin diffusion length for both HM and FM, spin transmission and reflection percentage at each interface, and THz emission efficiency.
\bibliography{aipsamp}

\end{document}